\documentclass{elsart}
\usepackage{graphicx}
\usepackage{amsmath}

\begin{document}

\begin{frontmatter}
\title{Localized patterns and hole solutions in one-dimension extended sytem}
\author{Marcel G. Clerc and Claudio Falcon}
\address{Departamento de F\'{\i}sica, Facultad de Ciencias
F\'{\i}sicas y Matem\'aticas, Universidad de Chile, Casilla 487-3,
Santiago, Chile.}

\begin{abstract}
The existence, stability properties, and bifurcation diagrams of
localized patterns and hole solutions in one-dimensional extended
systems is studied from the point of view of front interactions.
An adequate envelope equation is derived from a prototype model
that exhibits these particle-type solutions. This equation allow
us to obtain an analytical expression for the front interaction,
which is in good agreement with numerical simulations.
\end{abstract}

\end{frontmatter}

Non-equilibrium processes often lead in nature to the formation of
spatial periodic structures developed from a homogeneous state
through the spontaneous breaking of symmetries present in the
system \cite {Prigogine,Cross}. In the last decade localized
patterns or localized structures have been observed in different
experiments: liquid crystals \cite {Oswald}, gas discharge systems
\cite{Astrov97}, chemical reactions \cite {Swinney}, fluids
\cite{Fineberg}, granular media \cite{Melo}, and nonlinear optics
\cite{Arecchi,Ackemann}. One can understand these localized
patterns as patterns extend only over a small portion of the
spatial homogeneous systems. From the dynamic point of view,
localized patterns in one-dimensional spatial systems are the
homoclinic connection for the stationary dynamical system.
Recently, from a geometrical point of view, the existence,
stability properties, and bifurcation diagrams of localized
patterns in one-dimensional extended systems have been studied
\cite{Coullet2000}.

\begin{figure}[tbh]
\center
\includegraphics[width=13cm,height=6cm]{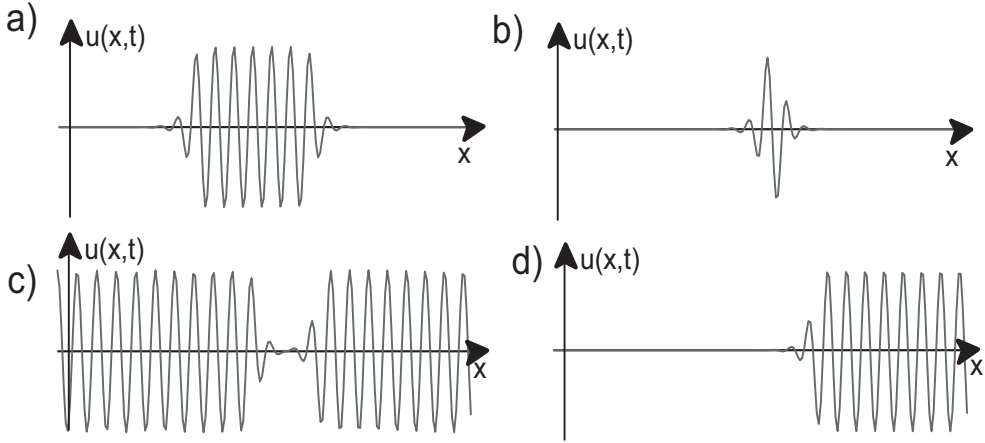}
\caption{Particle type solutions appear in the subcritical
Swift-Hohenberg equation. The parameters have been chosen as
$\epsilon=-0.16$, $\nu=1.00$, and $q=0.70$. (a) Localized pattern,
(b) shortest localized pattern, (c) hole solution and (d) front
solution.}
\end{figure}

The aim of this manuscript is to describe how one-dimensional
localized patterns and hole solutions arise from front
interactions. From a prototype model that exhibits localized
patterns and hole solutions, the subcritical Swift-Hohenberg
equation, we deduce an adequate equation for the envelope of these
particle type solutions. This model has a front solution that
connects a stable homogeneous state with an also stable but
spatially periodic one. Due to the oscillatory nature of the front
interaction, which alternates between attractive and repulsive, we
can infer the existence, stability properties, and bifurcation
diagrams of localized patterns and hole solutions. Hence, we
reobtain the bifurcation diagrams of localized patterns and hole
solution deduced from horseshoe behavior of the attractive and
repulsive manifold of ordinary differential equations
\cite{Coullet2000}.

Lets consider a prototype model that exhibits localized patterns
and hole solutions in one-dimension extended system, the
subcritical Swift-Hohenberg equation \cite {Brand96}:
\begin{equation}
\partial _{t}u=\varepsilon u+\nu u^{3}-u^{5}
-(\partial _{xx}+q^{2})^{2}u,
\label{E-SubcriticaSH}
\end{equation}
where $u\left( x,t\right) $ is an order parameter, $\varepsilon
-q^{4}$ is the bifurcation parameter, $q$ is the wave-number of
periodic spatial solutions, and $\nu $ is the control parameter of
the type of bifurcation, supercritical or subcritical. This model
describes the confluence of a stationary and an spatial
subcritical bifurcation, when the parameters scale as $u\sim
\varepsilon ^{1/4}$, $\nu \sim \varepsilon ^{1/2}$, $q\sim
\varepsilon ^{1/4}$, $\partial _{t}\sim \varepsilon $ and $
\partial _{x}\sim \varepsilon ^{1/4}$ ($\varepsilon \ll 1$). It
is often employed in the description of patterns observed in
Rayleigh-Benard convection and pattern forming systems \cite
{Cross}. In Fig. 1 we show typical localized patterns, hole
solutions, and motionless front solutions obtained by this model.
For small and negative $\nu$, and $9\nu ^{2}/40<\varepsilon <0$,
the system exhibits coexistence between a stable homogenous state
$u(x)=0$ and a periodic spatial one $ u(x)=\sqrt{\nu }\left(
\sqrt{2(1+\sqrt{1+40\varepsilon /9\nu })} \cos \left( qx\right)
\right) +o(\nu ^{5/2}) $. In this parameter region, one finds a
front between these two stable states (cf. Fig. 1). In order to
describe the front, localized patterns and hole solutions, we
introduce the ansatz
\begin{equation}
u=\sqrt{\frac{2\nu }{10}}\varepsilon ^{1/4}\left\{ A\left(
y=\frac{3\sqrt{ \left| \varepsilon \right| }}{2\sqrt{10q}}x,\tau
=\frac{9\nu ^{2}\left| \varepsilon \right| }{10}t\right)
+w_{1}\left( x,y,\tau \right) \right\} e^{iqx}+c.c,
\label{E-Ansatz}
\end{equation}
where $A(y,\tau )$ is the envelope of the front solution,
$w_{1}\left( x,y,\tau \right) $ is a small correction function of
order $\varepsilon $, and $\left\{ y,\tau \right\} $ are slow
variables. Note that in this ansatz we consider that $q$ is order
one, or larger that the other parameters. Introducing the above
ansatz in Eq. (\ref{E-SubcriticaSH}) and linearizing in $w_{1}$,
we find the following solvability condition
\begin{equation}
\partial _{\tau }A=\epsilon A+\left| A\right| ^{2}A-\left| A\right|
^{4}A+\partial _{yy}A+\left( \frac{A^{3}}{9\nu }-\frac{A^{3}\left|
A\right| ^{2}}{2}\right) e^{\frac{2iqy}{a\sqrt{\left| \varepsilon
\right| }}}-\frac{ A^{5}}{10}e^{\frac{4iqy}{a\sqrt{\left|
\varepsilon \right| }}}, \label{E-Envelope}
\end{equation}

where $\epsilon \equiv 10\varepsilon /9\nu ^{2}$, and $a\equiv
3\nu /2\sqrt{ 10}q$ . The terms proportional to the exponential
are non-resonant, that is, one can eliminate these terms by an
asymptotic change of variable. Furthermore, they have rapidly
varying oscillations in the limit $\epsilon \rightarrow 0$. Hence,
one usually neglects these terms. Note that, the above envelope
equation is a universal model, close to a spatial bifurcation, of
a system that exhibits coexistence between an homogeneous state
and spatial periodic one. In general, one can use an ansatz
similar to (\ref{E-Ansatz}) and noticing that the envelope
satisfies independently the symmetries $\left\{ x\rightarrow
-x,\text{ }A\rightarrow \bar{A}\right\} $, and $\left\{
x\rightarrow x+x_{o}\text{, }A\rightarrow Ae^{iqx_{o}}\right\}$
one derives equation (\ref{E-Envelope}).

When one considers only the resonant terms, it is straightforward
to show that the system has a front solution between two
homogeneous states, $0$ and $\left( 1+\sqrt{ 1+4\epsilon }\right)
/2$, when $-1/4<\epsilon <0$. This front propagates from the
global stable to the metastable one, and it is motionless when the
Maxwell point is reached at $\epsilon _{M}=-3/16$, and it has the
form
\begin{equation*}
a_{\pm }(y)=\sqrt{\frac{3/4}{1+e^{\pm \sqrt{3/4}(y-y_{o})}}}e^{i\theta },
\end{equation*}
where $y_{o}$ is the front's core position, and $\theta $ is an
arbitrary phase.

\begin{figure}[tbh]
\center
\includegraphics{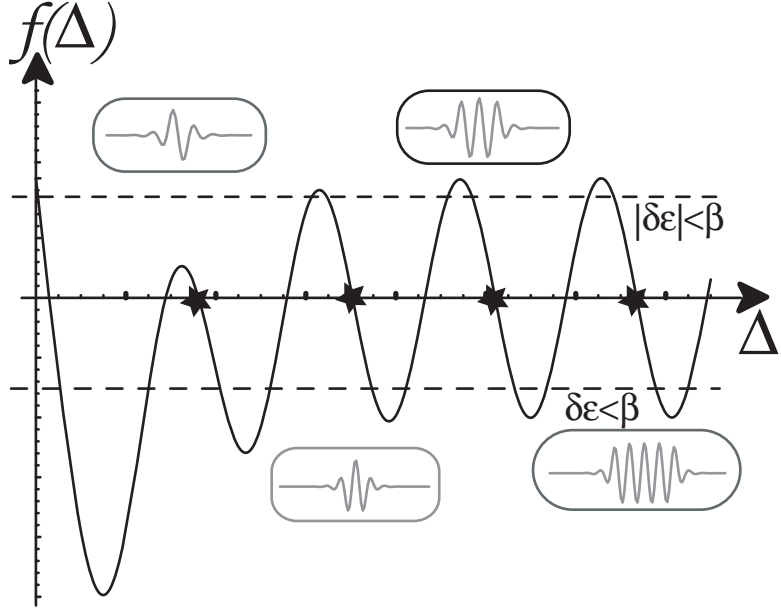}
\caption{Oscillatory interaction force  $f\left( \Delta \right)$.
The inset figures are the stable localized patterns observed at
the Maxwell point. The length of these localized patterns are
represented by the stars. The dashed lines represent the effective
abscissa that determine the size of localized patterns when
$\varepsilon$ is changed.}
\end{figure}

To describe a localized pattern exhibited by
(\ref{E-SubcriticaSH}) as a pair of two fronts, we must then
consider the non-resonant terms in the envelope equation
(\ref{E-Envelope}). We consider all these terms as perturbations
because they have rapidly varying oscillations. Close to the
Maxwell point, we use the ansatz
\begin{equation*}
A_{LP}(y,\tau)=\left[a_{-}(y-y_{1}(\tau))+a_{+}(y-y_{2}(\tau))-\rho
\left( y_{1},y_{2},y,\tau\right) \right]e^{i\theta \left(
y_{1},y_{2},y,\tau\right) },
\end{equation*}
where $\left\{ \rho ,\theta \right\} $ are small correction
functions, which are of order $\delta \epsilon \equiv (\epsilon
-\epsilon _{M})$ and $ y_{2}>y_{1}$. Introducing the above ansatz
in equation (\ref{E-Envelope}), linearizing in $\left\{ \rho
,\theta \right\} $ and after straightforward calculations, we
obtain the following solvability condition for the  distance
between the fronts
\begin{equation}
\frac{d\Delta }{d\tau}=f\left( \Delta \right) \equiv -\alpha
\Delta \exp (- \sqrt{\frac{3}{4}}\Delta )+\beta \cos (2q\Delta
/\sqrt{\epsilon })+2\delta \epsilon ,  \label{E-CoreFront}
\end{equation}
where $\Delta \equiv y_{2}-y_{1}$, $\alpha =27 \sqrt{3}/ 64$ and
$\beta =64\sqrt{3}q^{2}\exp (-q4\pi /\sqrt{\epsilon })/3\epsilon
$. In Fig. 2, we display the interaction of the front pair. It is
important to note that in one-dimensional extended systems, the
dependence of the distance in the front interaction is only
exponential \cite{Cross}. The linear  and periodic dependence of
$\Delta $ is a consequence of  the interaction (contained in the
non-resonant terms) of the large scale with the small scale
underlying the spatial periodic solution. The system has several
equilibria, $f(\Delta ^{\ast })=0$, and they are stable if $
f^{\prime }\left( \Delta^{\ast } \right) <0$. Thus, the existence
and stability of localized patterns is given by oscillatory nature
of the interaction. As is illustrated in Fig. 2, the region of
attractive and repulsive interaction is separated by localized
patterns. Note that, the larger equilibria represent localized
patterns with a bigger number of bumps. In order to understand the
bifurcation diagrams of localized patterns, we consider the effect
of change the bifurcation parameter $\epsilon $. Note that,
modifying $\epsilon $ is equivalent to move the abscissa of the
front interaction (cf. Fig. 2). First, we consider the case
$\left| \delta \epsilon \right| >\beta $ and $\delta \epsilon <0$;
the interaction is always attractive, that is, there is no
equilibrium. Hence, if one takes into account a front that
connects the homogenous state with spatial periodic state, then
the spatial periodic state invades the homogenous one.

In Fig. 3, the thick solid line is the velocity of propagation of
the front as function of the bifurcation parameter. Increasing
$\epsilon $, one finds the first equilibrium point $\Delta =\infty
$ for $\delta \epsilon =\delta \epsilon _{-}\equiv \beta $ and
$\delta \epsilon <0$. Thus the system has a motionless front
between the spatial periodic and homogenous states. Note that,
this equilibrium point remains meanwhile $\left| \delta \epsilon
\right| \leq \beta $, therefore this front is motionless in a
parameter range. This phenomenon is well-known as \textit{Locking
phenomenon} and the interval $\left| \delta \epsilon \right| \leq
\beta $ is denominated pining range \cite{Pomeau}. For $\delta
\epsilon>\beta $, the front propagates from the spatial periodic
state to the homogenous one. Increasing $\delta \epsilon $ from
$\delta \epsilon _{-},$ we observe that the equilibria, localized
patterns, appear by saddle-node bifurcation and with length
smaller than the previous ones, i.e., the localized patterns
appear by pair, one stable and another unstable, and each time
with less bumps. This sequence of bifurcations is illustrated in
Fig. 3 by the points $ c_{i}^{a}$. For $\delta \epsilon $ small,
and close to the Maxwell point, the system has a family of
infinite localized patterns. The length of the localized patterns
are roughly multiple of that of the shortest localized state.
Contrarily, for $\left| \delta \epsilon \right|
>\beta$, the localized patterns disappear by saddle node
bifurcation and increasing $ \delta \epsilon $ the larger
localized patterns disappear one after another. Hence, the
shortest localized state is the last to disappear. In Fig. 3 are
represent the sequence of these bifurcation by $c_{i}^{d}$.

\begin{figure}[tbh]
\center
\includegraphics{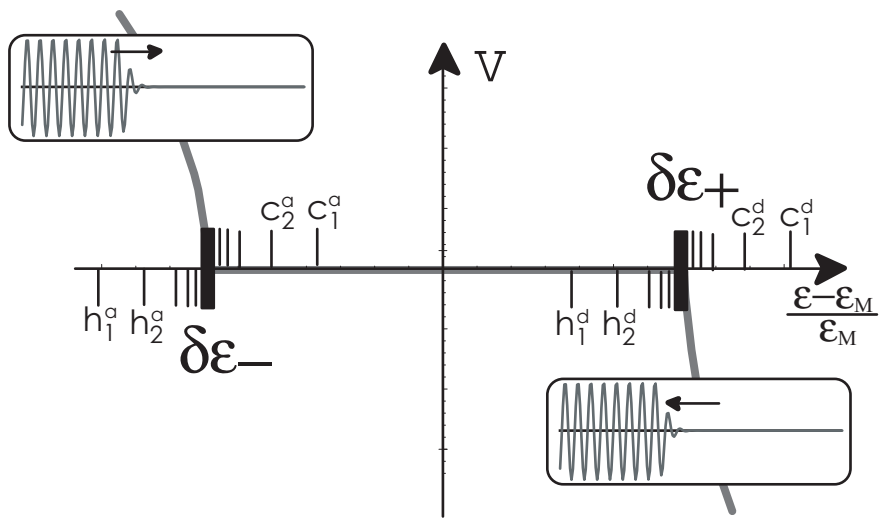}
\caption{Speed of the front and bifurcation diagrams of the
localized patterns and hole solutions as function of the
bifurcation parameter. The thick solid line is the analytical
formula of front speed. $c_{i}^{a}$ and $c_{i}^{d}$ ($h_{i}^{a}$
and $h_{i}^{d}$) represent the bifurcation points where the
localized patterns with (hole solution without) i-bumps appears
and disappear respectively.}
\end{figure}

The model (\ref{E-SubcriticaSH}) has different particle type
solutions as: front solution, localized patterns and hole
solutions. These solution are displayed in Fig. 1. From the front
interaction the hole solution can be understood as the
\textit{complement of localized patterns},  because these
solutions can be describe in term of the front as
\begin{equation*}
A_{Hole}(y,\tau)=(a_{+}(y-y_{2}(\tau))+a_{-}(y-y_{2}(\tau))-\rho
\left( y_{1},y_{2},y,\tau\right) )e^{i\theta \left(
y_{1},y_{2},y,\tau\right) },
\end{equation*}
where this solution asymptotically converges to a spatial periodic
state. We obtain the same expression for the interaction
(\ref{E-CoreFront}) changing $\alpha $ by $-\alpha $. Therefore
for these solutions, we obtain a similar bifurcation diagram of
localized patterns but inverted, that is, the first hole in to
appear and disappear by saddle-node bifurcation is the shortest
hole, and successively the hole with shorter length appears one
after the other and sequentially the solutions with shorter length
disappear one after the other. In Fig. 3 is illustrated this
sequence of bifurcations by $\left\{ h_{i}^{a},h_{i}^{a}\right\}
$. It is important to remark that the bifurcation diagram shown in
Fig. 3 have been deduced from geometrical arguments based in
horseshoe behavior of the attractive and repulsive manifold of a
ordinary differential equation \cite{Coullet2000}. In the pining
rage, the front solution is motionless. When one considers
additive white noise, the noise induces front propagation
\cite{Clerc2005}. The mean velocity of the front is zero only in
the Maxwell point, that is, at the Maxwell point the front core
describes a Brownian motion.

In conclusion, we have shown on the basis of the front
interactions the existence, stability properties, and bifurcation
diagrams of localized patterns and hole solutions in
one-dimensional extended systems.

The simulation software \textit{DimX} developed by P. Coullet and
collaborators at the laboratory INLN in France has been used for
all the numerical simulations. M.G.C. acknowledges the support of
FONDECYT project 1051117, FONDAP grant 11980002, and ECOS-CONICYT
collaboration  programs.


\begin{thebibliography}{99}
\bibitem{Prigogine}  G. Nicolis and I. Prigogine, \textsl{Self-Organization
in Non Equilibrium systems} (J.Wiley \& sons, New York, 1977).

\bibitem{Cross}  M. Cross and P. Hohenberg, Rev. Modern Phys. \textbf{65},
581 (1993).

\bibitem{Oswald}  S. Pirkl, P. Ribire and P. Oswald, Liq. Cryst. \textbf{13}
, 413 (1993).

\bibitem{Astrov97}  Y.A Astrov and Y.A. Logvin, Phys. Rev. Lett. \textbf{79}
, 2983 (1997).

\bibitem{Swinney}  K.L. Lee, W.D. McCormick, Q. Ouyang and H. Swinney,
Science \textbf{261}, 189 (1993).

\bibitem{Fineberg}  O. Liobashevski, Y. Hamiwl, A. Agnon, Z. Reches and J.
Fineberg, Phys. Rev. lett. \textbf{83}, 3190 (1999).

\bibitem{Melo}  Umbanhowar, F. Melo and H. Swinney, Nature \textbf{382}, 793
(1996).

\bibitem{Arecchi}  F.T. Arecchi, S. Boccaletti and P.L. Ramazza, Phys. Rep.
\textbf{318},1 (1999).

\bibitem{Ackemann}  B. Schapers, M. Feldmann, T. Ackemann and W. Lange,
Phys. Rev. Lett. \textbf{85}, 748 (2000).

\bibitem{Coullet2000}  P. Coullet, C. Riera and C. Tresser, Rev. Lett.
\textbf{84}, 3069 (2002).

\bibitem{Brand96}  H. Sakaguchi and H. Brand, Physica D 97, 274 (1996).

\bibitem{Pomeau}  Y. Pomeau, Physica D \textbf{23}, 3 (1986).

\bibitem{Clerc2005}  M.G. Clerc, C. Falcon and E. Tirapegui, submited to Phys. Rev. Lett.
\end{thebibliography}
\end{document}